\documentclass[article]{elsarticle}

\usepackage{lineno,hyperref}
\modulolinenumbers[5]

\journal{Nuclear Physics A}

\usepackage{graphicx} 
\usepackage{dcolumn} 
\usepackage{bm}         
\usepackage{amsmath}
\usepackage{slashed}
\usepackage{graphicx}
\usepackage{rotating}
\usepackage{float}

\newcommand{\ba}{\begin{eqnarray}}
\newcommand{\ea}{\end{eqnarray}}
\newcommand{\bd}{\begin{displaymath}}
\newcommand{\ed}{\end{displaymath}}
\newcommand{\Lagr}{\mathcal{L}}
\newcommand{\no}{\nonumber}

\def\bra#1{\left\langle #1\right|}
\def\ket#1{\left| #1\right\rangle}

\begin{document}

\begin{frontmatter}

\title{The chiral S=-1 meson-baryon interaction with new constrains on the NLO contributions}

\author{A. Ramos \fnref{myfootnote}, A. Feijoo and V.K. Magas}
\address{Departament de F\'isica Qu\`antica i Astrof\'isica and Institut de Ci\`encies del Cosmos, 
		       Universitat de Barcelona, Mart\'i i Franqu\`es 1, 08028 Barcelona, Spain}
\fntext[myfootnote]{Electronic address: ramos@ecm.ub.edu}




\begin{abstract}
We present a study of the $S=-1$ meson-baryon interaction, employing a chiral SU(3) Lagrangian up to next-to-leading order (NLO) and implementing unitarization in coupled channels. The parameters of the model have been fitted to a large set of experimental scattering data in different two-body channels, to threshold branching ratios, and to the precise SIDDHARTA value of the energy shift and width of kaonic hidrogen. In contrast to other groups, we have taken into consideration  the $K^- p\to K^+\Xi^-, K^0\Xi^0$ reaction data, since we found in a previous work to be  especially sensitive to the NLO parameters of the chiral Lagrangian. In the present work we also include the Born terms, which usually have very little effect, and find them to be non-negligible in the $K^- p\to K\Xi$ channels, correspondingly causing significant modifications to the NLO parameters.  We furthermore show that the importance of the Born terms becomes more visible in the isospin projected amplitudes of the $K^-p \to K\Xi$ reactions. The measurement of processes that filter single isospin components, like the $K^0_L p \to K^+ \Xi^0$ reaction that could be measured at the proposed secondary $K^0_L$ beam at Jlab, would
put valuable constraints on the chiral models describing the meson-baryon interaction in the $S=-1$ sector.
\end{abstract}

\begin{keyword}
$\bar{K}N$ interaction \sep NLO Chiral Lagrangian 
 \sep Coupled channel calculations \sep $\Xi$ production
\end{keyword}

\end{frontmatter}

\linenumbers

\section{Introduction}

It is well known that Quantum Chromodynamics (QCD) is strongly coupled at low energies and, thus, cannot be applied perturbatively to describe the interaction of hadrons in this regime. An appropriate framework is provided by effective theories, such as SU(3) Chiral Perturbation Theory ($\chi$PT), which respects the chiral symmetry of QCD or, more specifically, spontaneous chiral symmetry-breaking that causes the appearance of the Nambu-Goldstone (NG) bosons as light pseudoscalar mesons and the dynamical mass generation of hadrons \cite{Gasser:1983yg,Meissner:1993ah,Bernard:1995dp,Ecker:1994gg,Pich:1995bw}.
While $\chi$PT describes very satisfactorily hadron interactions at low energies, it fails in the vicinity of resonances, which are poles of the scattering amplitude, making the use of nonperturbative schemes mandatory. 

Unitarized Chiral Perturbation Theory (U$\chi$PT), which combines chiral dynamics with unitarization techniques in coupled channels, has shown to be a very powerful tool that permits extending the validity of $\chi$PT to higher energies and to describe the physics around certain resonances, the so called dynamically generated resonances (see \cite{ollerreport} and references therein). A clear example of the success of U$\chi$PT is  the description of the $\Lambda(1405)$ resonance, located only $27$ MeV below the $\bar{K} N$ threshold, that emerges from coupled-channel meson-baryon re-scattering in the $S=-1$ sector. In fact, the dynamical origin of the $\Lambda(1405)$ resonance was already hindered more than 50 years ago \cite{L1405}, an idea that was reformulated later in terms of the chiral unitary theory in coupled channels \cite{KSW}, and its success stimulated a lot of activity in the community \cite{OR,KWW,OM,LK,BMW,2pole,BFMS,BNW,BMN,GarciaRecio:2002td,GO,IHW,Mai:2012dt,Mizutani:2012gy}. The various developed models could reproduce the $\bar{K} N$ scattering data very satisfactorily and all these efforts culminated in establishing the $\Lambda(1405)$ as a superposition of two poles of the scattering amplitude  \cite{OM,2pole,PRL} generated dynamically from the unitarized meson-baryon interaction in coupled channels.

In this paper we present a study of the $S=-1$ meson-baryon interaction aiming to providing a well justified set of values for the low-energy constants of the next-to-leading order (NLO) chiral Lagrangian. To adjust our model parameters we will make use of experimental data on elastic and inelastic cross sections for $K^- p$ scattering  ($K^- p \to K^- p$, ${\bar K}^0 n$, $\pi^{\pm}\Sigma^{\mp}$, $\pi^0\Sigma^0$, $\pi^0\Lambda$),  the  threshold branching ratios \cite{br_1,br_2}, the precise SIDDHARTA value of the energy shift and width of kaonic hydrogen \cite{SIDD},  and, in contrast to what has commonly been done, we also employ the data from the reactions $K^- p\to K^+\Xi^-, K^0\Xi^0$. The motivation lies in the fact that
the lowest-order (LO) Lagrangian does not contribute directly to these reactions, which are then especially sensitive to the NLO terms as we have shown in our previous work \cite{Feijoo:2015yja}. The novelty of the present work is that we also consider the so called Born diagrams, which cause very little changes in all the studied channels, except the $K\Xi$ ones, as we will show. And, since these channels influence strongly the determination of the NLO parameters, their values might be significantly affected by the inclusion of the Born terms in the model. 

We also study the isospin decomposition of the $K\Xi$ production cross sections, and argue that the study of the meson-baryon interaction in $S=-1$ could highly benefit from isospin filtered reactions, such as the $K^0_L p \to  K^+ \Xi^0$ process, which could be measured in the recently proposed secondary $K^0_L$ beam at Jlab \cite{L_intent}. We present the predictions that our models, one from the present work and one from  \cite{Feijoo:2015yja}, give for this purely $I=1$ reaction.

\section{Chiral unitary approach}
\label{chiral}

\subsection{Lagrangian} 

This section provides a detailed development of the formalism employed for describing the meson-baryon scattering. We start with the SU(3) chiral effective Lagrangian which embeds the symmetries and chiral spontaneous symmetry breaking patterns of Quantum Chromodynamics,
\ba \label{LagrphiB1}
\Lagr_{\phi B}^{(1)} & = & i \langle \bar{B} \gamma_{\mu} [D^{\mu},B] \rangle
                            - M_0 \langle \bar{B}B \rangle 
                            - \frac{1}{2} D \langle \bar{B} \gamma_{\mu}
                             \gamma_5 \{u^{\mu},B\} \rangle \no \\
                     &   &  - \frac{1}{2} F \langle \bar{B} \gamma_{\mu} 
                               \gamma_5 [u^{\mu},B] \rangle \ ,
\ea
where we only present the term responsible for the interactions between mesons and baryons. Here the symbol $\langle \dots \rangle$ stands for the trace in flavor space, $M_0$ is the common baryon octet mass in the chiral limit, and the constants $D$, $F$ denote the axial vector couplings of the baryons to the mesons.
More specifically, this Lagrangian describes the coupling of the pseudoscalar octet $(\pi,K,\eta)$ to the fundamental baryon octet $(N,\Lambda,\Sigma,\Xi)$.

The pseudoscalar meson octet $\phi$ is arranged in a matrix valued field
\begin{equation}
\label{Uphi}
U(\phi) = u^2(\phi) = \exp{\left( \sqrt{2} i \frac{\phi}{f} \right)},
\end{equation} 
with
\begin{equation}
\label{mesfield}
\phi=\begin{pmatrix}
\frac{1}{\sqrt{2}}\pi^0+\frac{1}{\sqrt{6}}\eta & \pi^+ & K^+\\
\pi^- & -\frac{1}{\sqrt{2}}\pi^0+\frac{1}{\sqrt{6}}\eta & K^0\\
 K^- & \bar{K}^0 & -\frac{2}{\sqrt{6}}\eta\\
\end{pmatrix},
\end{equation} 
and $f$ being the pseudoscalar decay constant in the chiral limit. The quantity $U$ enters the  Lagrangian in the combination $u_\mu = i u^\dagger \partial_\mu U u^\dagger$.

The octet baryon fields are collected in
\begin{equation}
\label{Bfield}
B=\begin{pmatrix}
\frac{1}{\sqrt{2}}\Sigma^0+\frac{1}{\sqrt{6}}\Lambda & \Sigma^+ & p\\
\Sigma^- & -\frac{1}{\sqrt{2}}\Sigma^0+\frac{1}{\sqrt{6}}\Lambda & n\\
\Xi^- & \Xi^0 & -\frac{2}{\sqrt{6}}\Lambda\\
\end{pmatrix}  \ ,
\end{equation} 
and, finally, $[D_\mu, B]$ stands for the covariant derivative
\begin{equation}
\label{CoDer}
[D_\mu, B] = \partial_\mu B + [ \Gamma_\mu, B] \ ,
\end{equation} 
with $\Gamma_\mu =  [ u^\dagger,  \partial_\mu u] /2$ being the chiral connection. 
 
\begin{figure}[ht]
\begin{center}
 \centering
 \includegraphics[width=4.5in]{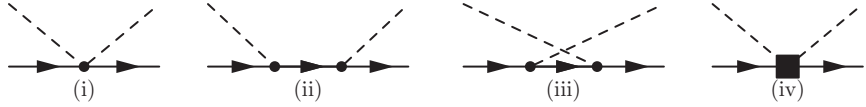}
\caption{Feynman diagrams for the meson-baryon interaction: Weinberg-Tomozawa term (i), direct and crossed Born terms (ii) and (iii), and NLO terms (iv). Dashed (solid) lines represent the pseudoscalar octet mesons (octet baryons). }
\label{Born}
\end{center}
\end{figure}

The relevant contributions to the interaction kernel are diagrammatically represented in Fig.~\ref{Born}. The  first three contributions appearing int the figure, namely (i), (ii) and (iii), are calculated using the Lagrangian at lowest order. 
%
%

The Weinberg-Tomozawa (WT) term corresponds to the  diagram (i) in Fig.~\ref{Born} and, in a non-relativistic limit, it reads: 
\begin{equation}
\label{WT}
V^{\scriptscriptstyle WT}_{ij}=-C_{ij}\frac{1}{4f^2} {\cal N}_i {\cal N}_j\left(\sqrt{s}-M_i-M_j\right),
\end{equation} 
where $\sqrt{s}$ is the total energy of the meson-baryon system in the center-of-mass frame. As we see, the WT term depends only on one parameter - the pion decay constant $f$, which is well known experimentally, $f_{\rm exp}=92.4$~MeV.  However in U$\chi$PT calculations this parameter is usually taken to be larger than the experimental value, ranging from $f=1.15 f_{\rm exp}$ to $f=1.36 f_{\rm exp}$ \cite{OR,KWW,OM,LK,BMW,2pole,BFMS,BNW,BMN,GarciaRecio:2002td,GO,IHW,Mai:2012dt,Mizutani:2012gy}, meaning to be a sort of average over the decay constants of the mesons involved in the various coupled channels.
We will leave it as a free parameter of our fits. 

The indices $(i,j)$ cover all the initial and final channels, which, in the case of strangeness $S=-1$ and charge $Q=0$ explored here, amount to ten: $K^-p$, $\bar{K}^0 n$, $\pi^0\Lambda$, $\pi^0\Sigma^0$, $\pi^-\Sigma^+$, $\pi^+\Sigma^-$, $\eta\Lambda$, $\eta\Sigma^0$, $K^+\Xi^-$ and $K^0\Xi^0$. The matrix of coefficients $C_{ij}$ can be found, for example, in Table VII of Ref. \cite{Feijoo:2015yja}. The normalization factor ${\cal N}$ is defined as ${\cal N}=\sqrt{(M+E)/(2M)}$, with $M$ and $E$ being, respectively, the mass and energy of the baryon. 

We next consider the s-wave projection and the non relativistic assumption for the the so called Born diagrams. The direct Born term, shown in Fig.~\ref{Born} (ii), is given by
\begin{eqnarray}
V^{D}_{ij} =
-\sum_{k=1}^{8}
\frac{C^{(\text{Born})}_{\bar{i}i,k}  C^{(\text{Born})}_{\bar{j}j,k}}{12 f^2}
\,{\cal N}_i\,{\cal N}_j\,
\frac{(\sqrt{s}-M_i)(\sqrt{s}-M_k)(\sqrt{s}-M_j)}{s-M_k^2}~~,
\label{eq:directBorn}
\end{eqnarray}
where the $k$ label refers to the intermediate baryon involved in the process. The coefficients $C^{(\text{Born})}_{\bar{i}i,k}$ can be found in Ref.~\cite{BNW} and they depend on the axial vector constants $D$ and $F$. In the same way, the crossed Born term, shown in Fig.~\ref{Born} (iii) , reads
\begin{align}
V^{C}_{ij} =&
\sum_{k=1}^{8}
\frac{C^{(\text{Born})}_{\bar{j}k,i} C^{(\text{Born})}_{\bar{i}k,j}}{12 f^2}\,
{\cal N}_i \,{\cal N}_j\ \nonumber\\
& \times
\biggl[
    \sqrt{s}+M_k
    -\frac{(M_i+M_k)(M_j+M_k)}{2 \left(M_i+E_i\right)\left(M_j+E_j\right)}
    (\sqrt{s}-M_k+M_i+M_j) \nonumber\\*
    &+\frac{(M_i+M_k)(M_j+M_k)}{4q_iq_j}
    \bigl\{
    \sqrt{s}+M_k-M_i-M_j \nonumber\\*
    &-\frac{s+M_k^2-m_i^2-m_j^2-2E_iE_j}{2\left(M_i+E_i\right)\left(M_j+E_j\right)}
    (\sqrt{s}-M_k+M_i+M_j)\bigr\} \nonumber \\
    &\times
    \ln\frac{s+M_k^2-m_i^2-m_j^2-2E_iE_j-2q_iq_j}
    {s+M_k^2-m_i^2-m_j^2-2E_iE_j+2q_iq_j}
    \biggr]~~\ ,
\label{eq:crossedBorn}
\end{align}
where $q_i,\ q_j$ are the center-of-mass (CM) three-momenta in the corresponding $i,j$ channels, and 
$m_i,m_j$ denote the corresponding meson masses.  
 
At NLO the contributions of $\Lagr_{\phi B}$ to meson-baryon scattering, shown by diagram (iv) in Fig.~\ref{Born},  are:
\begin{eqnarray}
    \Lagr_{\phi B}^{(2)}& = & b_D \langle \bar{B} \{\chi_+,B\} \rangle
                             + b_F \langle \bar{B} [\chi_+,B] \rangle
                             + b_0 \langle \bar{B} B \rangle \langle \chi_+ \rangle \no \\ 
                        &   &+ d_1 \langle \bar{B} \{u_{\mu},[u^{\mu},B]\} \rangle
                             + d_2 \langle \bar{B} [u_{\mu},[u^{\mu},B]] \rangle    \no \\ 
                        &   &+ d_3 \langle \bar{B} u_{\mu} \rangle \langle u^{\mu} B \rangle
                             + d_4 \langle \bar{B} B \rangle \langle u^{\mu} u_{\mu} \rangle \ ,
\label{LagrphiB2}
\end{eqnarray}
where $\chi_+ = 2 B_0 (u^\dagger \mathcal{M} u^\dagger + u \mathcal{M} u)$ breaks chiral symmetry explicitly via the quark mass matrix  $\mathcal{M} = {\rm diag}(m_u, m_d, m_s)$ and $B_0 = - \bra{0} \bar{q} q \ket{0} / f^2$ relates to the order parameter of spontaneously broken chiral symmetry.
As we see, there are 7 parameters at NLO, namely the low energy constants $b_D$, $b_F$, $b_0$ and $d_i$ $(i=1,\dots,4)$, which will be fitted to experimental data. 


From the Lagrangian $\Lagr_{\phi B}^{(2)}$ one can derive the meson-baryon interaction kernel at NLO:
\begin{equation}
\label{V_NLO}
V^{\scriptscriptstyle NLO}_{ij}=\frac{1}{f^2}{\cal N}_i {\cal N}_j\left[D_{ij}-2\left(\omega_i \omega_j+\frac{q_i^2q_j^2}{3\left(M_i+E_i\right)\left(M_j+E_j\right)}\right)L_{ij}\right],
\end{equation} 
where $\omega_i, \omega_j$ are the meson energies involved in the transition amplitude. The $D_{ij}$ and $L_{ij}$ coefficients depend on the NLO parameters $b_0$, $b_D$, $b_F$, $d_1$, $d_2$, $d_3$ and $d_4$ and are given, for example, in Table VIII of \cite{Feijoo:2015yja}.

The final interaction kernel employed in this work is expressed as the sum:
\begin{equation}
\label{V_TOT}
 V_{ij}=V^{\scriptscriptstyle WT}_{ij}+V^{D}_{ij}+V^{C}_{ij}+V^{\scriptscriptstyle NLO}_{ij}
\end{equation}

\subsection{Chiral coupled-channels dynamics}

It is well known that a perturbative treatment of the scattering amplitude fails in an energy region which contains resonances. Actually, in this sector of $S=-1$ and $Q=0$, the interaction between $\bar{K}$ mesons and nucleons is dominated by the presence of the $\Lambda(1405)$ resonance. Therefore, the correct treatment in this energy region requires the implementation of a non-perturbative resummation, for which we solve the 
Bethe-Salpether equation in coupled channels using the interaction kernel derived from the chiral Lagrangian,  Eq.~(\ref{V_TOT}). 

The Bethe-Salpetter method consists in solving a complex system of integral equations over the four-momentum of the intermediate  meson-baryon system in the CM fame. Even though this momentum might reach off-shell values, it has been shown \cite{OR,OO_ND,HJ_rev} that the off-shell parts of the interaction kernel gives rise to tadpole-type diagrams, which can be reabsorbed into renormalization of couplings and masses and could hence be omitted from the calculation. 
Thus, the scattering amplitude $T_{ij}$, which accounts for infinite contributions of the coupled channels, can be found solving the simple system of algebraic equations:
\begin{equation}
\label{LS_expanded}
T_{ij} =V_{ij}+V_{il} G_l V_{lj}+V_{il} G_l V_{lk}G_k V_{kj}+...\quad \Rightarrow  \quad T_{ij} = V_{ij} + V_{il} G_l T_{lj},
\end{equation} 
where the subscripts $i,j,l,\dots$ run over all possible channels and the loop function $G_l$ stands for the propagator of the meson-baryon state of channel $l$.

In the presence of the Born terms the above on-shell scheme should be treated with care since these may lead to conceptual and/or practical
difficulties, as described in Ref. \cite{Bruns:2010sv}. In particular, the u-channel Born graph, Fig. 1(iii), introduces unphysical subthreshold cuts, generated by the propagator of the intermediate baryon. In principle,  it is possible that the subthreshold cuts of some heavy meson-baryon pair can contribute to physical processes of light meson-baryon channels. This is an artifact of the on-shell scheme and a simple way to deal with it, consisting in matching the u-channel Born term to a constant value below a certain invariant energy,  was proposed in Ref. \cite{BNW}.  Fortunately, these unphysical cuts are not encountered in the kinematical regions explored in the present work.

The solution of Eq. (\ref{LS_expanded}) can be presented in matrix form as follows:
\begin{equation}
T ={(1-VG)}^{-1}V ,
 \label{T_algebraic}
\end{equation} 
where the loop function $G$ stands for a diagonal matrix with elements: 
\begin{equation} \label{Loop_integral}
G_l={\rm i}\int \frac{d^4q_l}{{(2\pi)}^4}\frac{2M_l}{{(P-q_l)}^2-M_l^2+{\rm i}\epsilon}\frac{1}{q_l^2-m_l^2+{\rm i}\epsilon} \ ,
\end{equation} 
with  $M_l$ and $m_l$ being the baryon and meson masses of the $l^{\rm th}$ channel. 
Applying the dimensional regularization method to deal with the logarithmic divergence of the above integral, one finds
\ba
& G_l = &\frac{2M_l}{(4\pi)^2} \Bigg \lbrace a_l+\ln\frac{M_l^2}{\mu^2}+\frac{m_l^2-M_l^2+s}{2s}\ln\frac{m_l^2}{M_l^2} + \no \\ 
 &     &\frac{q_{\rm cm}}{\sqrt{s}}\ln\left[\frac{(s+2\sqrt{s}q_{\rm cm})^2-(M_l^2-m_l^2)^2}{(s-2\sqrt{s}q_{\rm cm})^2-(M_l^2-m_l^2)^2}\right]\Bigg \rbrace \ .  
 \label{dim_reg}    
\ea 
The loop functions $G_l$ depend, for a given dimensional regularization scale $\mu$ which is taken to be $1$ GeV, on the subtraction constants $a_l$. These are unknown parameters that will be fitted to the experimental data. In fact, there are ten channels in this sector, but isospin symmetry reduces the subtraction constants to six independent ones. 

According to the normalization we use, the unpolarized total cross-section in the various meson-baryon scattering channels is given by:
\begin{equation}
\sigma_{ij}=\frac{1}{4\pi}\frac{M_iM_j}{s}\frac{k_j}{k_i}S_{ij} \ ,
 \label{sigma}
\end{equation}
where we have averaged over the initial baryon spin projections and summed over the final baryon spin projections:
\begin{equation}
\quad S_{ij}=\frac{1}{2}\sum_{s',s}|T_{ij}(s',s)|^2 \ .
 \label{M_matrix}
\end{equation}

Likewise, the $K^- p$ scattering length is obtained from the $K^- p$ scattering amplitude at threshold as:
\begin{equation}
a_{\scriptscriptstyle K^- p}=-\frac{1}{4\pi}\frac{M_p}{\sqrt{M_p+M_{\bar K}}}T_{K^- p \to K^- p}\ ,
 \label{scat_lenght}
\end{equation}
with
\begin{equation}
\quad T_{i \to j}=\frac{1}{2}\sum_{s',s} T_{ij}(s',s) \ .
 \label{T_matrix}
\end{equation}

The energy shift and width of the ground state of the kaonic hydrogen can be obtained from the scattering length $a_{\scriptscriptstyle K^- p}$ \cite{Meissner:2004jr}
by means of the second order corrected Deser-type formula:
\begin{equation}
\Delta E-i\frac{\Gamma}{2}=-2\alpha^3\mu_{r}^{2} a_{\scriptscriptstyle K^- p} \Big[ 1+2 a_{\scriptscriptstyle K^- p}\,\alpha\,\mu_r\, (1-\ln\alpha) \Big],                              \label{ener_shift_width}
\end{equation} 
where $\alpha$ is the fine-structure constant and $\mu_r$ the reduced mass of the $K^-p$ system.

One can also obtain the following measured ratios of elastic and inelastic $K^-p$ cross sections yields evaluated at threshold: 
\ba 
\gamma & = & \frac{\Gamma(K^- p \rightarrow \pi^+ \Sigma^-)}{\Gamma(K^- p \rightarrow \pi^- \Sigma^+)}  ,\\ \nonumber
 R_n & = & \frac{\Gamma(K^- p \rightarrow \pi^0 \Lambda)}{\Gamma(K^- p \rightarrow {\rm neutral\, states})}   , \\ \nonumber
 R_c & = & \frac{\Gamma(K^- p \rightarrow \pi^+ \Sigma^-,\pi^- \Sigma^+ )}{\Gamma(K^- p \rightarrow {\rm inelastic\, channels})} \ .     
\label{branch_ratios} 
\ea  

\section{Data treatment and fits}

This section is devoted to the discussion of the fitting procedure and the data treatment. In order to fix the parameters which appear in the different studied approaches, we consider a large amount of cross section data for $K^-p$ scattering into different final channels \cite{exp_1,exp_2}. Some points of these data sets have inconsistencies with the trend of the neighbouring points and have not been employed in the fitting procedure \footnote{There are only three such points and, therefore, we cannot give systematic criteria to exclude a point from the fit. We have excluded the points which produced a much higher $\chi^2$ than the neighbouring ones. These points were clearly not following the general trend of the corresponding data and fell completely outside of our error bands.}, leaving us with a total of 161 experimental points coming from $K^-p$ scattering. The eliminated points will be displayed in red in the figures. In addition, we take into account the recent energy shift and width of the 1s state of kaonic hydrogen obtained by the SIDDHARTA Collaboration \cite{SIDD}, namely $\Delta E_{1s}=283\pm36$ and $\Gamma_{1s}=541\pm92$, as well as the branching ratios defined in Eq.~(\ref{branch_ratios}) from Ref. \cite{br_1,br_2}. Summarizing, 166 points are included in the fit. 

The $\chi^2$ per degree of freedom used in this work is given by the expression:
\begin{equation}
\chi^{2}_{\rm d.o.f}=\frac{\sum_{k=1}^K n_k }{\left( \sum_{k=1}^K n_k -p \right)} \frac{1}{K} \sum_{k=1}^K \frac{\chi^{2}_{k}}{n_k}
\label{Chi^2_dof}
\end{equation}
with
\begin{equation}
 \chi^{2}_{k}=\sum_{i=1}^{n_k}\frac{\left( y_{i,k}^{\rm th}- y_{i,k}^{\rm exp}\right)^2}{\sigma_{i,k}^{2}} \ .\nonumber 
\end{equation}
With this definition we are using the method already adopted in \cite{GarciaRecio:2002td,GO,IHW,Mai:2012dt}, which consists in assigning an equal weight to each measurement when the total $\chi^2$ is calculated. This is achieved by averaging, over the number of observables $K$, the corresponding $\chi^2$ per degree of freedom,  which is obtained by dividing the total contribution of the experiment, $\chi^2_k$, by its own number of experimental points, $n_k$.
The number of free fitting parameters is denoted by $p$ . Such a  definition puts all the experiments on the same footing, since the relevance of those observables which have a small number of experimental points is not suppressed in favour of those with a larger set.

Three different fits have been performed in an attempt to learn about the importance of the different terms of the chiral Lagrangian. The first fit, called WT+Born(no $\Xi$), corresponds to a unitarized calculation employing the chiral Lagrangian at LO, that is, an interaction kernel which incorporates the contribution of the WT and the Born terms. This involves the fitting of nine parameters: the pion decay constant $f$, the six subtraction constants $a_{\bar{K}N}$,  $a_{\pi\Lambda}$, $a_{\pi\Sigma}$, $a_{\eta\Lambda}$, $a_{\eta\Sigma}$, and $a_{K\Xi}$, and the axial vector couplings of the baryons to the mesons $D$, $F$. The latter two are constrained to lie within 12.5\%  of their canonical value,  $D=0.8$ and $F=0.46$ \cite{Ratcliffe:1998su}. For this fit the experimental data from the $ K^- p \to K \Xi$ cross sections are excluded, as done in most of the works in the field, and we only take into account 91 points which are compiled in \cite{br_1,br_2,SIDD,exp_1}. 

The second fit, called WT+NLO+Born(no $\Xi$) is the natural extension of the previous one, where the NLO terms of the interaction kernel are incorporated, thus involving seven additional parameters: the NLO low energy constants $b_0$, $b_D$, $b_F$, $d_1$, $d_2$, $d_3$ and $d_4$. Similar fits have been performed in \cite{BNW,GO,IHW,Mizutani:2012gy} to which we will compare our results. 

It was demonstrated in \cite{Feijoo:2015yja} that the $K^- p \to K \Xi$ channels are the most sensitive to the NLO terms in the chiral Lagrangian, and therefore the $\Xi$ production data was essential to constraint the NLO fits. Having this in mind,  we present results for a third fit, called WT+NLO+Born, which is similar to the WT+NLO+Born(no $\Xi$) one, but also employs the $K\Xi$ production cross section data \cite{exp_2} in the $\chi^{2}_{\rm d.o.f}$ minimization. This fit complements those in the study of \cite{Feijoo:2015yja}, by taking into account the Born terms, which were not accounted for there. 

\begin{table*}[!ht]
\begin{tabular}{lcccc}
\hline \\[-2.5mm]
                               &  WT                      & WT+NLO          &  WT+NLO     & WT+NLO \\
                               & +Born(no $\Xi$)      & +Born(no $\Xi$) &       +Born     & \cite{Feijoo:2015yja} \\
\hline \\[-2.5mm]
$a_{\bar{K}N} \ (10^{-3})$     & $-1.45\pm 1.59$      & $0.16\pm 0.38$   & $1.77\pm 2.38$ & $6.55 \pm 0.63$ \\
$a_{\pi\Lambda}\ (10^{-3})$    & $-12.7\pm 9.5$       & $31.7\pm 1.3$    & $55.2\pm 13.5$ & $ 54.8 \pm 7.5$\\
$a_{\pi\Sigma}\ (10^{-3})$     & $-0.11\pm 4.32$      & $-2.04\pm 0.93$  & $2.33\pm 3.17$ & $ -2.29 \pm 1.89$\\
$a_{\eta\Lambda}\ (10^{-3})$   & $-1.49\pm 3.65$      & $76.0\pm 1.3$    & $8.00\pm 5.04$ & $ -14.2 \pm 12.7$ \\
$a_{\eta\Sigma}\ (10^{-3})$    & $1.0\pm 11.4$        & $-18.3\pm 1.3$   & $6.5\pm 20.6$  & $ -5.17 \pm 0.07$\\
$a_{K\Xi}\ (10^{-3})$          & $23.1\pm 85.0$       & $-2.57\pm 0.14$  & $-9.04\pm 3.63$ & $ 27.0 \pm 7.8$ \\
$f/f_{\pi}$                    & $1.22\pm 0.04$       & $1.18\pm 0.03$   & $1.21\pm 0.03$  & $ 1.20 \pm 0.01$\\
$b_0 \ (GeV^{-1}) $            & \multicolumn{1}{c}{-}  & $-0.71\pm 0.01$  & $-0.70\pm 0.23$ & $ -1.21 \pm 0.01$\\
$b_D \ (GeV^{-1}) $            & \multicolumn{1}{c}{-}  & $0.47\pm 0.01$   & $0.31\pm 0.20$ & $ 0.05 \pm 0.04$\\
$b_F \ (GeV^{-1}) $            & \multicolumn{1}{c}{-}  & $0.50\pm 0.02$   & $0.65\pm 0.41$ & $ 0.26 \pm 0.15$\\
$d_1 \ (GeV^{-1}) $            & \multicolumn{1}{c}{-}  & $0.42\pm 0.06$   & $0.17\pm 0.26$ & $ -0.11 \pm 0.06$\\
$d_2 \ (GeV^{-1}) $            & \multicolumn{1}{c}{-}  & $0.20\pm 0.01$   & $0.17\pm 0.11$ & $ 0.65 \pm 0.02$\\
$d_3 \ (GeV^{-1}) $            & \multicolumn{1}{c}{-}  & $0.22\pm 0.03$   & $0.37\pm 0.16$ & $ 2.85 \pm 0.04$\\
$d_4 \ (GeV^{-1}) $            & \multicolumn{1}{c}{-}  & $-0.21\pm 0.02$  & $0.01\pm 0.09$ & $ -2.10 \pm 0.02$\\
$D$                            & $0.70\pm 0.16$       & $0.90\pm 0.13$   & $0.90\pm 0.10$  & \multicolumn{1}{c}{-}\\
$F$                            & $0.41\pm 0.06$       & $0.51\pm 0.08$   & $0.40\pm 0.08$ & \multicolumn{1}{c}{-}\\
\hline \\[-2.5mm]
$\chi^2_{\rm d.o.f.}$          & $0.56$                 & $0.41$             & $0.73$           &  $0.65$      \\
\hline
\end{tabular}
\caption{Values of the parameters and the corresponding  $\chi^2_{\rm d.o.f.}$, defined as in Eq.~(\ref{Chi^2_dof}), for the different fits described in the text. The value of the pion decay constant is $f_{\pi}=92.4$ MeV and the subtraction constants are taken at a regularization scale $\mu=1$ GeV. The NLO fit performed in \cite{Feijoo:2015yja} has been included in the last column of the table, denoted here as WT+NLO to be consistent with the labelling of the other fits. The error bars of the parameters are those given by the MINUIT minimization procedure.} 
\label{tab2}
\end{table*}


We have also estimated the error bands for the $K^-p$ scattering cross sections into different final meson-baryon channels for our best fit. We have followed the method employed in Ref. \cite{GO} for our definition of $\chi^2$, Eq. (\ref{Chi^2_dof}), which is proposed in an earlier study \cite{Lampton:1976an}.
First of all, we calculated the {\it correlated} error bars for our model parameters, by generating new parameter configurations by randomly varying all the free parameters around their central values through a Monte Carlo generator, and rejecting those configurations for which the corresponding value of  $\chi^2$ (total) satisfies
\begin{equation}
\chi^2>\chi_{0}^2+\chi^2(p,1\sigma)\,, 
\label{chi_cond}
\end{equation}
where $\chi_{0}^2$ corresponds to the minimum found by MINUIT and $\chi^2(p,1\sigma)$ is the value of a chi-squared distribution with $p$ d.o.f. at a confidence level of one sigma. 
In the next step, we generated 16.000 of new parameter configurations by randomly varying all the free parameters within the obtained correlated error bars, and the $K^-p$ scattering cross sections obtained for these configurations determine their corresponding error bands. Similarly, we can also associate error bars to the threshold observables from the values obtained with these new configurations.

\section{Results and discussion}

\subsection{Influence of the Born terms}

In a previous work \cite{Feijoo:2015yja}, 
we studied the meson-baryon interaction in the $S=-1, Q=0$ sector,  
aiming for a better comprehension of the relevance and the role played by the terms next in hierarchy after the WT one. We focused on the  $K \Xi$ production reactions because they do not receive direct lowest order contributions and the rescattering terms are insufficient to reproduce the experimental data properly. It was indeed found that the NLO terms were crucial to obtain a good reproduction of the $\bar{K}N \to K^+ \Xi^-, K^0 \Xi^0$ cross sections.  Subsequently, we included, only for $\bar{K}N \to K \Xi$ reactions, high spin and high mass resonances applying the Rarita-Schwinger method as it was done in \cite{Nakayama:2006ty,Man:2011np,Shyam:2011ys,Sharov:2011xq,Jackson:2015dva}. From the eight possible candidates present in this energy range \cite{PDG}, and according to the findings of the resonant model of \cite{Sharov:2011xq},  we chose the $\Sigma(2030)$ and $\Sigma(2250)$ resonances as best candidates. A better agreement with the experimental data was achieved, together with a  considerable improvement of the accuracy and stability of the NLO coefficients.

This work was carried out under the assumption that, at lowest order, the contribution of the $s$ and $u$-channel diagrams involving the coupling of the meson-baryon channel to an intermediate baryon state, i.e. the Born terms (ii) and (iii) of Fig. \ref{Born},  would be very moderate \cite{OM}. The idea was also reinforced by other works, namely \cite{BNW,Mizutani:2012gy}. In these articles, among other approaches, the authors present two models based on chiral Lagrangians which contain WT and NLO terms and where the difference between the models was the inclusion or not of the Born terms. Concretely, "Model c" and "Model s" from \cite{Mizutani:2012gy}, and fits "c" and "u" from \cite{BNW}. As a result, one can see that the inclusion of the Born terms led to rather small changes in the fitting parameters and did not influence significantly the quality of the fits, although in some cases it helped in producing more natural values of the subtracting constants. 
A similar behavior when the Born diagrams are included is observed in \cite{IHW}.   

\begin{table*}[!ht]
\begin{tabular}{lccc}
\hline \\[-2.5mm]
                      &  {$a_p(K^-p \rightarrow K^- p)$} & {$\Delta E_{1s}$} & {$\Gamma_{1s}$} \\
\hline \\[-2.5mm]
WT+Born(no $\Xi$)     & $-0.71+{\rm i\,}0.81$ & 307 & 572 \\%
WT+NLO+Born(no $\Xi$) & $-0.66+{\rm i\,}0.80$ & 284 & 536 \\%
WT+NLO+Born           & $-0.67+{\rm i\,}0.84$ & 291 & 558\\%
                      & $(^{+0.17}_{-0.21})+ {\rm i\,}(^{+0.14}_{-0.17})$ &	$^{+76}_{-57}$ & $^{+92}_{-128}$ \\
\hline \\[-2.5mm]
Exp.                  & $ -0.66 + {\rm i\,}0.81$ &	$283$ & $541$ \\
                      & $(\pm0.07)+ {\rm i\,}(\pm0.15)$ &	$\pm36$ & $\pm92$ \\
\hline
\end{tabular}
\caption{Threshold observables obtained from the WT+Born(no $\Xi$), WT+NLO+Born(no $\Xi$) and WT+NLO+Born fits explained in the text. The corresponding error bars for the WT+NLO+Born fit, obtained as described in the text, are also given. Experimental data are taken from \cite{SIDD}. }
\label{tab:thresh_1}
\end{table*}

\begin{table*}[!ht]
\begin{tabular}{lccc}
\hline \\[-2.5mm]
                     & {$\gamma$} & {$R_n$} & {$R_c$}  \\
\hline \\[-2.5mm]
WT+Born(no $\Xi$)    & 2.36 & 0.191 & 0.664  \\%
WT+NLO+Born(no $\Xi$)& 2.38 & 0.190 & 0.664  \\%
WT+NLO+Born          & 2.36 & 0.191 & 0.664 \\%
               & $^{+1.19}_{-0.74}$ & $^{+0.082}_{-0.075}$ & $^{+0.018}_{-0.011}$ \\
\hline \\[-2.5mm]
Exp.           & $2.36$ &	$0.189$ & $0.664$  \\
               & $\pm 0.04$ & $\pm 0.015$ & $\pm 0.011$ \\
\hline
\end{tabular}
\caption{Threshold observables obtained from the WT+Born(no $\Xi$), WT+NLO+Born(no $\Xi$) and WT+NLO+Born fits explained in the text. The corresponding error bars for the WT+NLO+Born fit, obtained as described in the text, are also given. Experimental data are taken from \cite{br_1,br_2}. }
\label{tab:thresh_2}
\end{table*}

\begin{figure}[h!]
 \centering
 \includegraphics[width=4.2in]{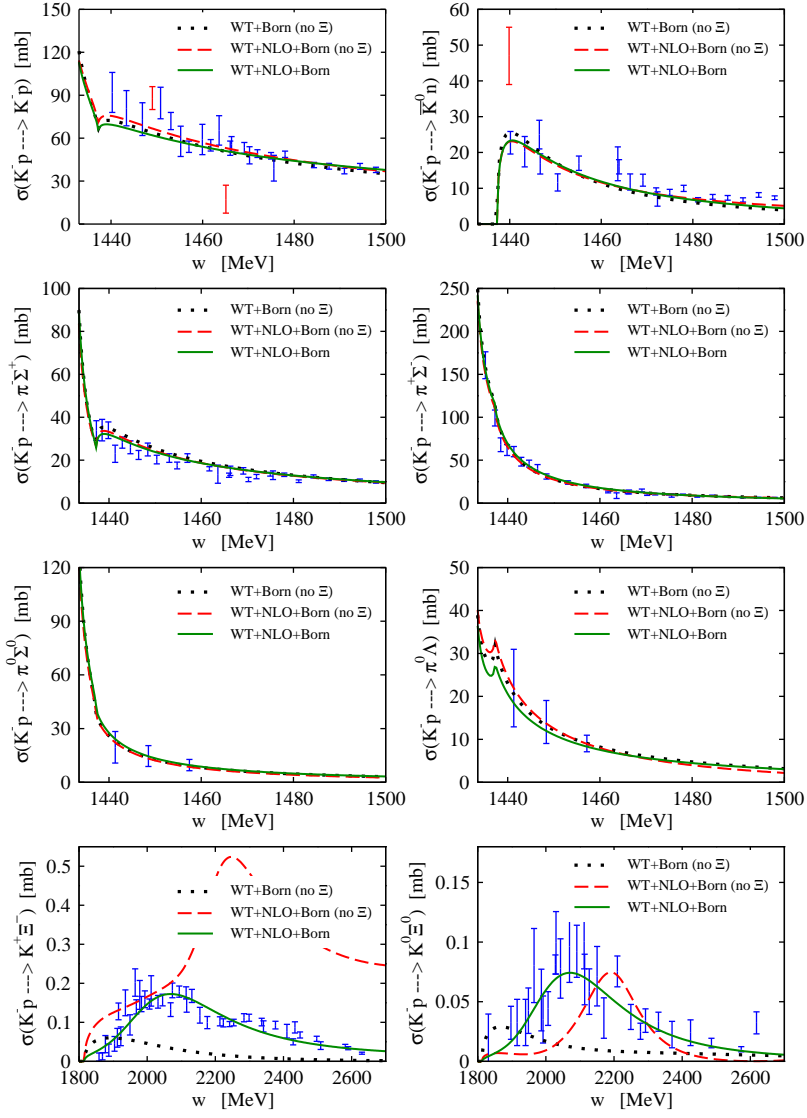}
 \caption{Total cross sections for the $K^- p\to K^- p$, $\bar{K}^0n$, $\pi^- \Sigma^+$, $\pi^+\Sigma^-$, $\pi^0 \Sigma^0$, $\pi^0\Lambda$, $K^0 \Xi^0$, $K^+ \Xi^- $ reactions obtained from the WT+Born(no $K\Xi$) fit (dotted line), the WT+NLO+Born(no $K\Xi$) fit (dashed line) and the WT+NLO+Born fit (solid line), where the last case take into account the experimental data of the $K\Xi$ channels, see text for more details. Experimental data are from \cite{exp_1,exp_2}. The points in red have not been included in the fitting procedure.}
 \label{Born_terms_channels}
\end{figure}

\begin{figure}[h!]
 \centering
 \includegraphics[width=4.2in]{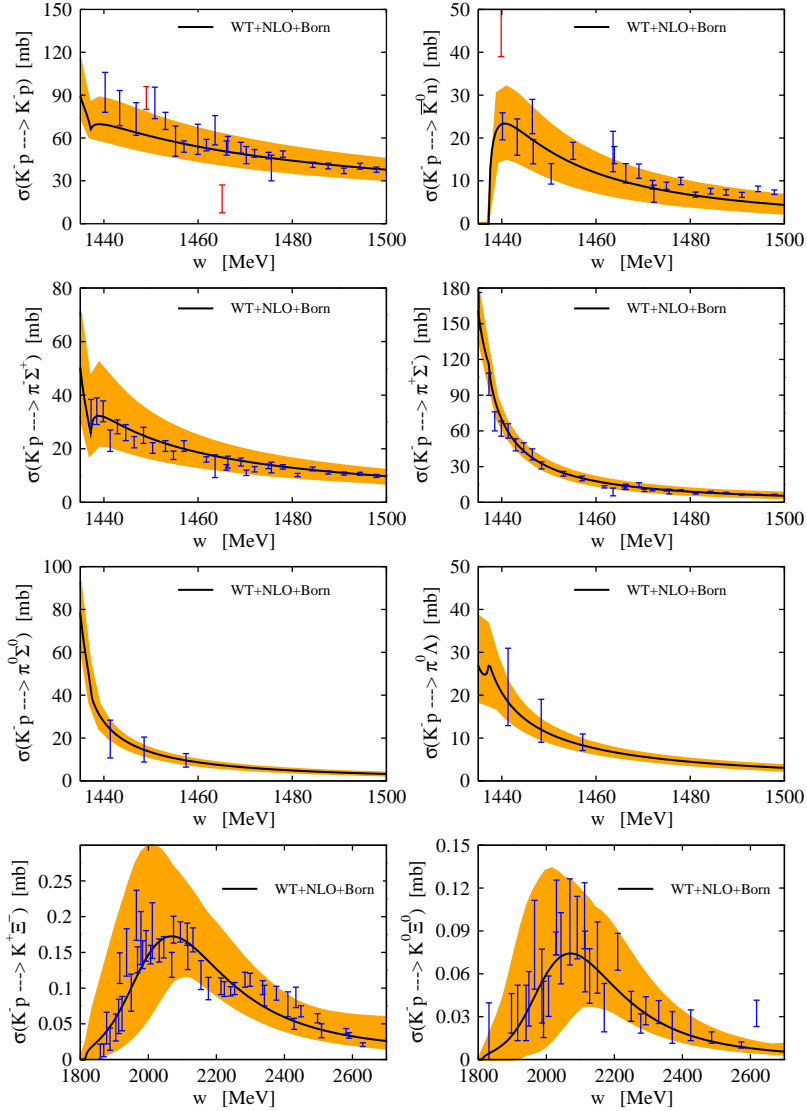}
 \caption{Total cross sections for the $K^- p\to K^- p$, $\bar{K}^0n$, $\pi^- \Sigma^+$, $\pi^+\Sigma^-$, $\pi^0 \Sigma^0$, $\pi^0\Lambda$, $K^0 \Xi^0$, $K^+ \Xi^- $ reactions obtained for the WT+NLO+Born fit (solid line), with our estimation of the corresponding error bars (yellow area), see text for more details. 
 Experimental data are from \cite{exp_1,exp_2}. The points in red have not been included in the fitting procedure.}
 \label{error_band}
\end{figure}

We next present the results of our fits when the Born terms are included as  described in the previous section. The model parameters obtained in these fits and the final $\chi^{2}_{\rm d.o.f}$ are summarized in Table \ref{tab2}. The results for the observables are shown in Tables~\ref{tab:thresh_1}, \ref{tab:thresh_2} and Figs. \ref{Born_terms_channels}, \ref{error_band}. As one can see, the overall agreement with the experimental data is very good. We recall that two of our fits, those denoted by the '(no $\Xi$)' label, have not included the $ K^+ \Xi^-, K^0 \Xi^0$ data. 

Fig. \ref{error_band}  presents our estimation of the error bands of the $K^-p$ scattering cross section into different final channels for our best fit, WT+NLO+Born, that gives $\chi_{0}^2=0.73\left( \sum_{k=1}^K n_k -p \right)$. In our particular case we have  $\sum_{k=1}^K n_k=166$, $p=16$, $\chi^2(16,1\sigma)=18.07$, and thus the rejection condition, Eq. (\ref{chi_cond}), leads to $\chi^{2}_{\rm d.o.f}>0.85$.

From our results, we first note that, comparing the WT+Born(no $\Xi$) and WT+NLO+Born(no $\Xi$) fits, one sees that the inclusion of the NLO terms do help improving the agreement with experimental data, as was shown by all the works that have used the chiral Lagrangian up to NLO.  The  '(no $\Xi$)' fits give some strength in the $K^+ \Xi^-,  K^0 \Xi^0$ channels, but do not reproduce the observed structure. Only if the $K\Xi$ production data is included into the fit we can get a reasonable agreement - see WT+NLO+Born results in Tables \ref{tab2}, \ref{tab:thresh_1}, \ref{tab:thresh_2} and Fig. \ref{Born_terms_channels}. We note that in this latter fit the $D$ and $F$ parameters of the Born contributions reached the edges of their allowed range, but in a way that their sum stays rather close to its nominal value $g_A=D+F=1.26$ \cite{Ratcliffe:1998su}.  

\begin{figure}[h!]
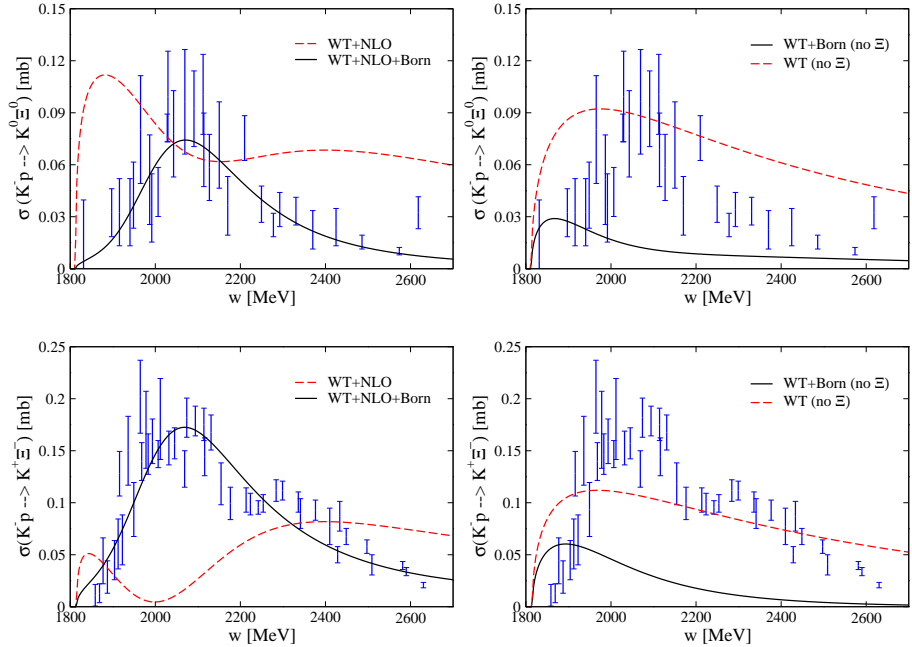

 \centering
 \includegraphics[width=0.49\textwidth]{Born_justification.eps}
 \includegraphics[width=0.49\textwidth]{Born_justification_2.eps}
 \caption{ 
 {\it Left column:\ } Total cross sections for the $K^- p\to K^0 \Xi^0, K^+ \Xi^- $ reactions obtained from 
the WT+NLO+Born fit (solid line), and the similar result neglecting the contributions from the Born terms (dashed line), see text and Fig. \ref{Born_terms_channels} for more details. 
Experimental data are from \cite{exp_2}. The points in red have not been included in the fitting procedure. 
{\it Right column:\ } The same as the left column, but the solid line correspond to WT+Born fit, where the presented  $K^- p\to K^0 \Xi^0, K^+ \Xi^- $ were not taken into account, see text and Fig. \ref{Born_terms_channels} for more details. The dashed line shows the similar result neglecting the contributions from the Born terms.   
}
 \label{Born_terms_justification}
\end{figure}

The most notable fact of the obtained  NLO parameters is that most of them acquire very different values from what was found in  \cite{Feijoo:2015yja}, as shown in Table~\ref{tab2}, although the only difference of the present model is the inclusion of the Born terms. The other effects induced by the Born terms are pretty similar to what was seen in \cite{BNW,Mizutani:2012gy,IHW}: the goodness of our WT+NLO+Born  fit is almost the same as that of the corresponding fit in \cite{Feijoo:2015yja}, and the  values of the subtraction constants are much closer to the 'natural size'  compared to the ones in \cite{Feijoo:2015yja} (Table III), in spite of the substantial associated errors. 
Another remarkable and expected result is that the inclusion or no inclusion of the $K\Xi$ production data does not affect the NLO parameters of our full model (WT+NLO+Born) fits  as strongly as was observed in \cite{Feijoo:2015yja}. In other words, although the $\chi^2$ is a bit worse when we take into account the Born contributions, these work as an important stabilizer for the NLO parameters. 

Although a more exhaustive analysis might be needed to fully understand their behavior, it is clear that the Born terms have a much stronger effect in our model than we expected, based on the study of other groups.  Since the only significant difference of our model with respect to those of \cite{BNW,Mizutani:2012gy,IHW} is the inclusion of the $\bar{K}N \to K \Xi$ data, we decided to investigate how important are the Born terms in these reactions.
 
For this purpose, we display in Fig.~\ref{Born_terms_justification} (left column)  our results of the  WT+NLO+Born fit (solid line), together with the results of the same fit but neglecting the contributions from the Born diagrams in Eq. (\ref{V_TOT}) (dashed line). As one can see,  including or not the Born terms leads to substantial changes in the $K\Xi$ production cross sections, meaning that these terms contribute at the same order of the other terms in the Lagrangian with which they can interfere strongly. 
The above statement is reconfirmed by the right column in Fig.~\ref{Born_terms_justification}, where we presented the WT+Born(no $\Xi$) fit (solid line), together with the results of the same fit but neglecting the contributions from the Born diagrams. These graphs clearly show that the importance of the Born terms for the $K \Xi$ production reactions is not related to the inclusion of the NLO terms. In other words, for these particular channels the strength coming from WT term, Born terms and NLO terms are of the same order, and should all be taken into account.

The relevance of the Born diagrams of the chiral model in the $\bar{K}N \to K^+\Xi^-, K^0 \Xi^0$ cross sections may not come as a surprise if one considers that a recent work, studying these reactions from a phenomenological  resonance model  \cite{Jackson:2015dva}, also finds substantial contributions coming from the exchange of the ground state $1/2^+$ hyperons in s- and u-channel exchange configurations.


\subsection{Isospin decomposition}

\begin{figure}[h!]
 \centering
 \includegraphics[width=3.2in]{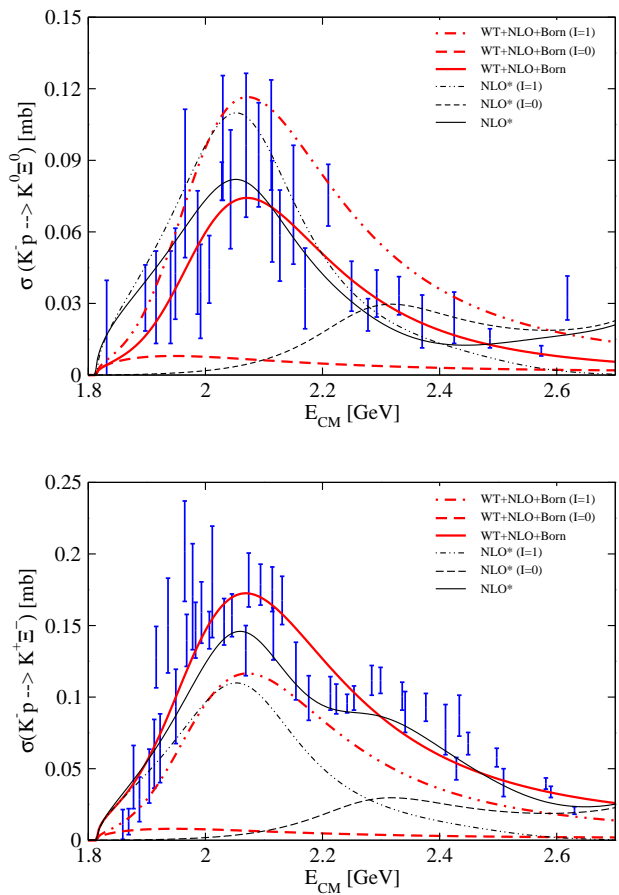}
 \caption{The total cross sections of the $K^- p\to K^0 \Xi^0$ reaction (top figure) and the $K^- p\to K^+ \Xi^-$ reaction (bottom figure) for the WT+NLO+Born model (thick lines) and for the NLO* model from Ref. \cite{Feijoo:2015yja} (thin lines), see text for more details. The solid lines show the results of the full amplitude, while the dashed and dash-dotted lines denote the $I=0$ and $I=1$ contributions respectively. Experimental data are from \cite{exp_2}.}
 \label{kXi_born_iso}
\end{figure}

In Fig.~\ref{kXi_born_iso} we split the contributions to the $K^- p\to K^0 \Xi^0,  K^+ \Xi^-$ cross sections coming from the different isospin components of the ${\bar K}N \to K\Xi$ amplitudes, according to:
\begin{eqnarray}
\langle K^-  p  \mid T \mid  K^+ \Xi^-\rangle &=& ~\frac{1}{2}\left[
 \langle \bar{K}  N \mid T^{I=1} \mid  K  \Xi \rangle 
- \langle \bar{K}  N  \mid T^{I=0} \mid  K  \Xi \rangle  \right] \\
\langle K^- p  \mid T \mid K^0 \Xi^0\rangle &=& -\frac{1}{2}\left[
 \langle \bar{K}  N  \mid T^{I=1} \mid  K  \Xi \rangle+
\langle \bar{K}  N  \mid T^{I=0} \mid  K  \Xi \rangle 
\right]\,.
\end{eqnarray}
The solid lines show the results obtained with the full amplitudes, while the dashed and double-dot dashed correspond to the results employing only the $I=0$ and $I=1$ components, respectively. 
We see that,
as a consequence of the inclusion of the Born terms, the isospin decomposition of the scattering amplitudes of the present model (thick lines) is distributed in a quite different way than in the models of \cite{Feijoo:2015yja} (thin lines), where these terms are neglected. As one can see, the $I=1$ component of our previous 
NLO$^*$ model (see more details in \cite{Feijoo:2015yja}) peaks at $2050$ MeV, while the $I=0$ component is concentrated at higher energies reaching its maximum at around $2300$ MeV. This is not the case for the our new fit which takes into account the Born terms. The $I=1$ component becomes much more prominent at higher energies and its peak is shifted $50$ MeV upwards. 
Conversely, the $I=0$ component presents a much lower strength and is located towards substantially smaller energies, peaking at around $1900$ MeV.

The fact that different models present such different isospin decompositions 
points towards the need of identifying reactions that proceed through either $I=0$ or $I=1$, thereby acting as isospin filters that help constraining the parameters of the meson-baryon lagrangian better. One example is 
the scattering into $\eta \Lambda$ final states, which are of pure isospin $0$. The data is already available \cite{exp_3}  an we plan to include it in a future fitting procedure in order to find more reliable values of the NLO coefficients. In this respect we would also like to mention the weak $\Lambda_b$ decay into a $J/\Psi$ and a meson-baryon pair, a reaction that filters the $I=0$ component in the final state, as it was shown in \cite{Roca:2015tea,Feijoo:2015cca}. These decays are presently analyzed   by the CDF \cite{Aaltonen:2010pj} and LHCb \cite{Aaij:2013oha, Aaij:2014zoa, Aaij:2015tga} collaborations. In particular, the  $\Lambda_b\to J/\Psi~K^-~ p$ decay has been employed very recently in \cite{Aaij:2015tga} to claim the presence of an exotic pentaquark charmonium state in the $J/\Psi\, p$ channel.

\subsection{$K^0_L p\to  K^+ \Xi^0$ reaction}

\begin{figure}[h!]
 \centering
 \includegraphics[width=3.2in]{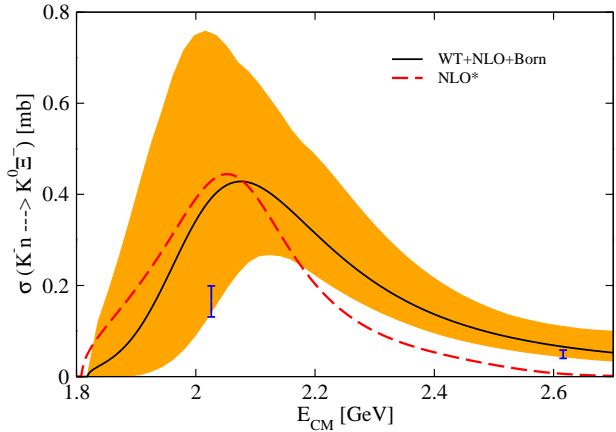}
 \caption{Total cross sections of the $K^0_L p\to  K^+ \Xi^0$ reactions for the for the 
 WT+NLO+Born model (full line) and for the NLO* model from Ref. \cite{Feijoo:2015yja} (dashed line), and the experimental points of the $I=1$ $K^- n \to  K^0 \Xi^-$ reaction, taken from  \cite{iso1exp1,iso1exp2} and divided by two, see text for more details. The error band corresponds to that of the WT+NLO+Born model.
}
 \label{kXi_Q=1_S=-1}
\end{figure}

It would also be interesting to obtain information on the ${\bar K}N \to K \Xi$ interacion in $I=1$. At the moment, only two data points, obtained from $K^-$ deuteron reactions in bubble chamber experiments \cite{iso1exp1,iso1exp2} are known for this component. The recent proposal \cite{L_intent} of creating a secondary $K^0_L$ beam at Jlab offers a great opportunity for measuring the $K^0_L p \to K^+ \Xi^0$ reaction. Since $K^0_L = (K^0 -{\bar K}^0)/\sqrt{2}$, the former reaction would proceed through the ${\bar K}^0$ component of the $K^0_L$, and, thus, would be of pure $I=1$ character. 
Our predictions for this reaction are shown in Fig.~\ref{kXi_Q=1_S=-1} for the  WT+NLO+Born model and for the NLO* model from Ref. \cite{Feijoo:2015yja}, together with the experimental points of the pure $I=1$ $K^- n \to  K^0 \Xi^-$ reaction, which have been divided by $2$ to properly account for the size of the strangeness $S=-1$ component of the $K^0_L$. 

We would like to remind the reader that these two data points have not been used in our fitting neither in this work nor in Ref. \cite{Feijoo:2015yja}. As one can see, our full model prediction does a good job at higher energy, but both predictions overshoot the lower energy point around $2$ GeV by a factor of 2, although it is worth mentioning that it falls within the error band of the WT+NLO+Born model. New data from the proposed secondary $K^0_L$ beam at Jlab would certainly be most welcome to help constraining the theoretical models.

\section{Conclusions}
We have presented a study of the $S=-1$ meson-baryon interaction, employing a chiral SU(3) Lagrangian up to NLO and implementing unitarization in coupled channels. The parameters of the model have been fitted to a large set of experimental scattering data in different two-body channels, to threshold branching ratios, and to the precise SIDDHARTA value of the energy shift and width of kaonic hidrogen. In contrast to other works, we have also constrained our model to also reproduce the $K^- p\to K^+\Xi^-, K^0\Xi^0$ reactions, and in this study we reconfirm the finding of Ref. \cite{Feijoo:2015yja} that these channels are especially sensitive to the NLO parameters of the chiral Lagrangian. Furthermore, we have shown that the Born terms, which usually have very little effect, become non-negligible in the $K^- p\to K\Xi$ channels, and correspondingly might lead to significant changes of the NLO parameters obtained in the fit.  

Most of the data employed in our fits are coming from antikaon proton scattering and  therefore contain contributions from both isospin $I=0$ and $I=1$ components; the only exception is the $\pi^0 \Sigma^0$ channel, which selects $I=0$. However, we have shown that inclusion or non inclusion of the Born terms as well as inclusion or non inclusion of the high spin resonances, which was studied in Ref. \cite{Feijoo:2015yja}, can seriously modify the isospin decomposition of the $K^- p\to K\Xi$ reactions.   
This is why data filtering $I=0$ or $I=1$ components  would be extremely valuable to further constrain the parameters of the chiral lagrangian describing the meson-baryon interaction in the $S=-1$ sector.

One such opportunity arises from the weak decay of the $\Lambda_b$ into states containing a $J/\Psi$ and meson-baryon pairs, measured by the CDF \cite{Aaltonen:2010pj} and LHCb \cite{Aaij:2013oha, Aaij:2014zoa, Aaij:2015tga} collaborations. A recent theoretical study of Ref. \cite{Feijoo:2015cca} has shown that such a reaction filters the final meson-baryon components in $I=0$. 

On the other hand, measuring the $K^0_L p \to K^+ \Xi^0$ reactions in $I=1$ with a secondary $K^0_L$ beam at Jlab \cite{L_intent} would be most welcome as a complement of the information one can obtain from $K^- p$ scattering data.  We have presented our prediction for this reaction based on our current best fit as well as on the one from \cite{Feijoo:2015yja}. 

We consider these results as preliminar, since we suspect that the inclusion of data coming from the scattering of $ \eta \Sigma^0, \eta \Lambda$ could be helpful to constrain the fitting parameters and to improve their precision. It is also posible that the consideration of the high-spin resonaces in  $K^- p \to K \Xi$ channels, as it was done in Ref. \cite{Feijoo:2015yja}, will lead to modification of the NLO parameters. We hope to address these questions in forthcoming works.

\section{Acknowledgements}

This work is partly supported by the Spanish Ministerio de Economia y Competitividad (MINECO) under the project MDM-2014-0369 of ICCUB (Unidad de Excelencia 'Mar\'\i a de Maeztu'), 
and, with additional European FEDER funds, under the contract FIS2014-54762-P, 
by the Ge\-ne\-ra\-li\-tat de Catalunya contract 2014SGR-401,
and 
by the Spanish Excellence Network on Hadronic Physics FIS2014-57026-REDT.

\section{References}


\end{document}